\documentclass[twocolumn,showpacs,showkeys,amsmath,amssymb]{revtex4}

\usepackage{graphicx}

\begin{document}

\title{Time Delay in Black Hole Gravitational Lensing as a Distance Estimator}

\author{V. Bozza$^{a,b,c}$, L. Mancini$^{b,c,d}$}
 \affiliation{$^a$ Centro studi e ricerche ``Enrico Fermi'', Rome, Italy. \\
 $^b$ Dipartimento di Fisica ``E. R. Caianiello", Universit\`a di
Salerno, 84081 Baronissi, Italy. \\
 $^c$ Istituto Nazionale di Fisica Nucleare, Sezione di Napoli, Italy. \\
 $^d$ Institut f\"{u}r Theoretische Physik der
           Universit\"{a}t Z\"{u}rich, CH-8057 Z\"{u}rich, Switzerland.}%

\date{\today}

\begin{abstract}
We calculate the time delay between different relativistic images
formed by black hole gravitational lensing in the strong field
limit. For spherically symmetric black holes, it turns out that
the time delay between the first two images is proportional to the
minimum impact angle. Their ratio gives a very interesting and
precise measure of the distance of the black hole. Moreover, using
also the separation between the images and their luminosity ratio,
it is possible to extract the mass of the black hole. The time
delay for the black hole at the center of our Galaxy is just few
minutes, but for supermassive black holes with $M=10^8 \div 10^9
\mathrm{M}_\odot$ in the neighbourhood of the Local Group the time
delay amounts to few days, thus being measurable with a good
accuracy.
\end{abstract}



\maketitle

\section{Introduction}
Gravitational lensing is a useful tool to investigate a lot of
aspects of the nature of the universe. It was the first prove of
the validity of the theory of general relativity (GR) \cite{DED},
and today its effects on extragalactic scales (lensing of quasars,
arcs in galaxy clusters, etc.) and on galactic scales
(microlensing) are ordinarily observed and studied by the
scientific community in the weak field approximation \cite{SEF}.

In the last years, a new form of gravitational lensing has been
proposed as a method to investigate the gravitational field
generated by collapsed objects. This approach considers light rays
of background sources passing very close to the event horizons of
black holes without entering inside. The study of this extreme
case is of remarkable interest: on one hand it represents an
independent test of GR in strong gravitational fields; on the
other hand, intrinsic features of the lens (rotation, electric
charge, etc.) could become accessible to the observations, opening
a new possibility to constrain black hole models.

In Schwarzschild framework, a light ray with small impact
parameter can wind several times around a black hole without being
definitively caught inside. In this way, a set of infinite
relativistic images can be generated on each side of the black
hole \cite{Atk,VirEll}. Several approaches have been proposed to
study gravitational lensing in the strong field limit
\cite{VirEll,Vie,Nem,FN,FKN,Per,DabSch}, while Falcke, Melia \&
Agol, in a different perspective, studied the accretion flow as a
source \cite{FMA}. In Ref. \cite{BCIS} analytical formulae for the
position and the magnification of the images were obtained,
defining a {\it strong field limit} for the deflection angle.
These formulae were applied to a Reissner-Nordstrom black hole in
Ref. \cite{ERT}, and were also used to calculate relativistic
effects on microlensing events \cite{Pet}. A full generalization
of the strong field limit for any spherically symmetric spacetime
was drawn in Ref. \cite{Boz1} and applied to several black hole
metrics, allowing a non-degenerate discrimination among different
collapsed objects. In work \cite{Bha}, the method was used to
examine the characteristics of a
Gibbons-Maeda-Garfinkle-Horowitz-Strominger charged black hole of
heterotic string theory. In work \cite{Boz2} the case of a Kerr
black hole and the relevance of its spin in strong field lensing
approximation was discussed for light rays travelling on
quasi-equatorial trajectories. Waiting for an analytical treatment
including non-equatorial trajectories, the general case is
explored numerically in \cite{VazEst}.

When multiple images are formed, the light-travel-time along light
paths corresponding to different images is generally not the same.
So, if the source is characterized by luminosity variations, these
variations would show up in the images with a relative temporal
phase depending on the geometry of the lens \cite{Ref}. These time
delays are usually measured in gravitational lensing observations
on cosmological scales. The striking importance of time delay lies
in the fact that it is the only dimensional observable. Therefore
its measurement is useful to determine at first the length scale
for a gravitational lensing system and its mass. Measuring the
time delays in cosmological contexts, it is possible to determine
the cosmological distance scale and hence the Hubble parameter
\cite{Ref,BlaNar,WCW}. This fact has drawn a great attention by
the scientific community towards this kind of measurements.

In the present paper, we estimate the time delay between images
generated by strong field lensing of black holes. We show that
time delays between relativistic images are indeed measurable in
most supermassive black holes suitable for gravitational lensing
studies. Moreover, it turns out that in a first approximation the
time delay between consecutive relativistic images is proportional
to the minimum impact angle. The ratio between these two
observables is nothing but the distance of the lens, that can be
estimated in a very precise way and without bias. Combining all
information, it is also possible to get an independent mass
estimate, to be compared with estimates obtained by other methods.

This paper is structured as follows. In Sect. 2, we recall the
main results of the strong field limit method. In Sect. 3 we
derive a general expression for the time delay, specifying it to
the spherically symmetric case. In Sect 4, we estimate the
expected time delays for several interesting supermassive
extragalactic black holes discussing the whole information that
can be extracted from a time delay measurement. Finally, in Sect.
6, we draw the conclusions. An appendix contains the computation
of the time delay for the Kerr metric, as an example of a
non-spherically symmetric metric.

\section{The strong field limit approach}
The technique we use in the derivation of the time delay between
different images resembles the main calculation of the deflection
angle in the strong field limit approach. We shall briefly recall
the main steps of that derivation referring the reader to Refs.
\cite{Boz1,Boz2} for all the details.

Consider a generic black hole metric projected on the equatorial
plane
\begin{equation}
ds^2=A(x)dt^2-B(x)dx^2-C(x)d\phi^2 +D(x) dt d\phi
\end{equation}
where $D(x)$ can be consistently set to zero in a spherically
symmetric black hole.

The metric does not depend on time and the azimuthal angle $\phi$,
so that, for a photon moving in this background, $\dot t$ and
$\dot \phi$ can be expressed in terms of two integrals of motion,
namely energy and angular momentum. By a suitable choice of the
affine parameter, we set the first to 1 and the second to the
impact parameter $u$ of the incoming photon. We have

\begin{eqnarray}
&& \dot t = \frac{4C-2u D}{4AC+D^2} \label{tdot} \\%
&& \dot \phi = \frac{4A u+2D}{4AC+D^2}. \label{phidot}
\end{eqnarray}

The impact parameter $u$, is related to the closest approach
distance $x_0$ by
\begin{equation}
u =\frac{-D_0+\sqrt{4A_0 C_0 +D_0^2}}{2A_0},
\end{equation}
where all functions with the subscript 0 are evaluated for
$x=x_0$.

By the on-shell condition for the photon, we also derive
\begin{equation}
\dot x = \pm \frac{2}{\sqrt{B}} \sqrt{\frac{C-u D-u^2A}{4AC+D^2}}.
\label{xdot}
\end{equation}

Dividing Eq. (\ref{phidot}) by Eq. (\ref{xdot}), we get

\begin{eqnarray}
&& \frac{d\phi}{dx}=P_1(x,x_0)P_2(x,x_0) \label{dphidx}\\%
&& P_1(x,x_0)=\frac{\sqrt{B}(2A_0 A u+ A_0 D)}{\sqrt{C A_0}
\sqrt{4AC+D^2}} \\%
&& P_2(x,x_0)=\frac{1}{\sqrt{A_0-A \frac{C_0}{C}+\frac{u}{C}(A
D_0- A_0 D )}}. \label{P2}
\end{eqnarray}

Integrating this expression from $x_0$ to infinity we find half
the deflection angle as a function of the closest approach. Given
the symmetry between approach and departure, we can write the
whole deflection angle as
\begin{eqnarray}
&& \alpha(x_0)=\phi_f(x_0)-\pi \\ %
&& \phi_f(x_0)=2\int\limits_{x_0}^\infty \frac{d\phi}{dx} dx.
\label{phif}
\end{eqnarray}

To solve this integral, we define the variables
\begin{eqnarray}
&& y=A(x) \\%
&& z= \frac{y-y_0}{1-y_0}
\end{eqnarray}
where $y_0 \equiv A_0$. The integral (\ref{phif}) in the
deflection angle becomes
\begin{eqnarray}
&& \phi_f(x_0)=\int\limits_0^1 R(z,x_0) f(z,x_0) dz \label{I z} \\%
&& R(z,x_0)=2\frac{1-y_0}{A'(x)}P_1(x,x_0) \label{R} \\%
&& f(z,x_0)=P_2(x,x_0) \label{f}
\end{eqnarray}
where $x=A^{-1} \left[\left(1-y_0 \right) z+ y_0 \right]$.

The function $R(z,x_0)$ is regular for all values of $z$ and
$x_0$, while $f(z,x_0)$ diverges for $z \rightarrow 0$. We then
expand the argument of the square root in $f(z,x_0)$ to the second
order in $z$, defining
\begin{equation}
 f(z,x_0) \sim f_0(z,x_0)= \frac{1}{\sqrt{\alpha z +\beta z^2}}.
 \label{f0}
\end{equation}

The Eq. $\alpha=0$ defines the radius of the photon sphere $x_m$,
which is the minimum approach distance for photons not falling
into the black hole.

The result of the integral (\ref{I z}) gives the strong field
limit expansion of the deflection angle \cite{Boz1}

\begin{equation}
\alpha(u)=-\overline{a} \log \left(\frac{u}{u_m}-1 \right)
+\overline{b} +O\left(u-u_m \right), \label{S F L theta}
\end{equation}
where the coefficients of the expansion are
\begin{eqnarray}
&& u_m=\frac{-D_m+\sqrt{4A_m C_m +D_m^2}}{2A_m} \label{um} \\
&& \overline{a}= \frac{R(0,x_m)}{2\sqrt{\beta_m}} \label{ob1}\\%
&& \overline{b}=-\pi+b_D+b_R+\overline{a} \log \frac{c x_m^2}{u_m}
\label{ob2}
\end{eqnarray}
and
\begin{eqnarray}
&&\!\! \!\!\!\! \!\!\!\!\!\! \!\!\!\! \!\!\!\! b_D=2 \overline{a}
\log \frac{2(1-y_m)}{A'_m x_m}
\\ && \! \!\!\!\! \!\!\!\! \!\! \!\!\!\! \!\!\!\!b_R=\int\limits_0^1\left[
R(z,x_m)f(z,x_m)-R(0,x_m)f_0(z,x_m) \right]dz,
\end{eqnarray}
while $c$ is defined by the expansion
\begin{equation}
u-u_m=c \left(x_0-x_m \right)^2. \label{xtou}
\end{equation}
All the functions with the subscript $m$ are evaluated at
$x_0=x_m$.

With the formula (\ref{S F L theta}) for the deflection angle, it
is straightforward to calculate the positions and the
magnifications of all relativistic images. Two infinite patterns
of relativistic images appear on each side of the lens, very close
to the minimum impact angle $\theta_m=u_m/D_{OL}$ ($D_{OL}$ is the
distance of the lens from the observer). These images are highly
demagnified unless the source is very close to a caustic point.
For spherically symmetric black holes, all caustic points are
exactly aligned with the lens, so that a source aligned with the
optical axis (the line joining observer and lens) would enhance
the magnification of all images simultaneously.

In spinning black holes, the caustics drift away from the optical
axis, so that one source cannot be simultaneously close to
different caustics. In this case only one image at a time can be
enhanced while all others stay very faint \cite{Boz2}.
Nevertheless, in this case, additional images, appearing when the
source is inside a caustic, may play an important role in the
phenomenology, yet to be understood.

For later reference, we write here the formula for the position of
the relativistic images
\begin{equation}
\theta^{\pm}_n=\pm \theta_m \left(1+ e^{\frac{\overline{b}-2n
\pi\pm \gamma}{\overline{a}}} \right). \label{theta}
\end{equation}
Here $\gamma$ is the angular separation between the source and the
optical axis, as seen from the lens. $n$ is the number of loops
done by the photon around the black hole. For each $n$, we have an
image on each side of the lens, according to the chosen sign.

\section{Time delay in the strong field limit} \label{Sec TD}
In this section we derive the time delay between different
relativistic images, following an approach similar to the one
reported in the previous subsection for the deflection angle, but
with some tricky subtraction strategies to treat the integrals.

For an observer at infinity, the time taken from the photon to
travel from the source to the observer is simply

\begin{equation}
T=\int\limits_{t_0}^{t_f} dt.
\end{equation}

Changing the integration variable from $t$ to $x$, we split the
integral into approach and leaving phases
\begin{equation}
T=\int\limits_{D_{LS}}^{x_0} \frac{dt}{dx}  dx+
\int\limits^{D_{OL}}_{x_0} \frac{dt}{dx}  dx.
\end{equation}

Here $D_{LS}$ is the distance between the source and the lens,
while $D_{OL}$ is the distance between the lens and the observer.

Extending the integration limits to infinity, we can unify the two
integrals into one, exploiting the symmetry between approach and
departure. This can be done at the price of subtracting two terms
\begin{equation}
T=2\int\limits_{x_0}^{\infty} \left| \frac{dt}{dx} \right|dx-
\int\limits_{D_{OL}}^{\infty} \left| \frac{dt}{dx} \right|dx-
\int\limits_{D_{LS}}^{\infty} \left| \frac{dt}{dx} \right|dx.
\end{equation}

If we consider two photons, travelling on different trajectories,
the time delay between them is
\begin{eqnarray}
&T_1 &-T_2 =2\int\limits_{x_{0,1}}^{\infty} \left|
\frac{dt}{dx}(x,x_{0,1}) \right|
dx-2\int\limits_{x_{0,2}}^{\infty} \left| \frac{dt}{dx}(x,x_{0,2})
\right|  dx \nonumber \\ &&-\int\limits_{D_{OL}}^{\infty} \left|
\frac{dt}{dx}(x,x_{0,1}) \right| dx+\int\limits_{D_{OL}}^{\infty}
\left|\frac{dt}{dx}(x,x_{0,2})\right| dx \nonumber \\
&&-\int\limits_{D_{LS}}^{\infty} \left|
\frac{dt}{dx}(x,x_{0,1})\right| dx+\int\limits_{D_{LS}}^{\infty}
\left| \frac{dt}{dx}(x,x_{0,2})\right| dx. \label{T1T2}
\end{eqnarray}

Supposing that observer and source are very far from the black
hole, $dt/dx$ is effectively 1 in the last four integrals which
thus exactly cancel each other. We are thus left with the first
two integrals.

Dividing Eq. (\ref{tdot}) by Eq. (\ref{xdot}), we obtain
\begin{eqnarray}
&&\frac{dt}{dx}=\tilde P_1(x,x_0)P_2(x,x_0) \\%
&&\tilde P_1(x,x_0)= \frac{\sqrt{B A_0}(2 C- u D)}{\sqrt{C }
\sqrt{4AC+D^2}} \label{P1t}
\end{eqnarray}
and $P_2$ defined by Eq. (\ref{P2}). Of course, $dt/dx$ tends to
one for large $x$ and the two integrals in (\ref{T1T2}) are
separately divergent, while their difference is finite. In fact,
the time delay is the result of the different paths followed by
the photons while they wind around the black hole. When the two
photons are far away from the black hole, $dt/dx \rightarrow 1$
and the two integrals compensate each other. Separating the two
regimes, we can write individually convergent integrals. To
achieve this, we subtract and add the function $\tilde
P_1(x,x_{0,i})/\sqrt{A_{0,i}}$ to each integrand. Supposing
$x_{0,1}<x_{0,2}$, we can write
\begin{eqnarray}
&&T_1-T_2= \tilde T(x_{0,1})-\tilde T(x_{0,2})
+2\int\limits^{x_{0,2}}_{x_{0,1}}
\frac{\tilde P_1(x,x_{0,1})}{\sqrt{A_{0,1}}} dx \nonumber \\ %
&&+ 2\int\limits_{x_{0,2}}^{\infty} \left[\frac{\tilde
P_1(x,x_{0,1})}{\sqrt{A_{0,1}}}- \frac{\tilde
P_1(x,x_{0,2})}{\sqrt{A_{0,2}}} \right] dx \label{T1T2b}
\end{eqnarray}
with
\begin{eqnarray}
&& \!\!\!\!\!\!\!\!\!\! \tilde T(x_0)=\int\limits_0^1 \tilde R(z,x_0) f(z,x_0) dz \label{Tt} \\%
&& \!\!\!\!\!\!\!\!\!\!  \tilde R(z,x_0)=2\frac{1-y_0}{A'(x)}
\tilde P_1(x,x_0) \left(1-\frac{1}{\sqrt{A_0}f(z,x_0)} \right)
\label{Rtilde}
\end{eqnarray}
and $f(x,x_0)$ defined by Eq. (\ref{f}). Substituting all the
expressions back into (\ref{T1T2b}), we can check that it is
equivalent to (\ref{T1T2}), but now it is written as a sum of
separately convergent integrals.

In practice, the integral $\tilde T(x_0)$ represents the time
spent by the light ray to wind around the black hole. In order to
cutoff the integrands at large $x$'s, in the definition of
$R(z,x_0)$ we have subtracted a term which is negligible when the
photon is close to the black hole but cancels the integrand when
the photon is far from the black hole. The residual terms of this
subtraction are stored in the last two integrals in (\ref{T1T2b})
and are generally subleading with respect to $\Delta \tilde T$, as
we shall see later.

The integral (\ref{Tt}) can be solved following the same technique
of the integral (\ref{I z}) in the previous subsection, just
replacing $R$ by $\tilde R$. The result is

\begin{equation}
\tilde T(u)=-\tilde{a} \log \left(\frac{u}{u_m}-1 \right)
+\tilde{b} +O\left(u-u_m \right) \label{Tt u}
\end{equation}
where $u_m$ is defined by Eq. (\ref{um}) and
\begin{eqnarray}
&& \tilde{a}= \frac{\tilde R(0,x_m)}{2\sqrt{\beta_m}} \label{at}\\%
&& \tilde{b}=-\pi+\tilde b_D+\tilde b_R+\tilde{a} \log \frac{c
x_m^2}{u_m} \label{bt}
\end{eqnarray}
with
\begin{eqnarray}
&&\!\! \!\!\!\! \!\!\!\!\!\! \!\!\!\! \!\!\!\! \tilde b_D=2
\tilde{a} \log \frac{2(1-y_m)}{A'_m x_m}
\\ && \! \!\!\!\! \!\!\!\! \!\! \!\!\!\! \!\!\!\! \tilde b_R=\int\limits_0^1\left[
\tilde R(z,x_m)f(z,x_m)-\tilde R(0,x_m)f_0(z,x_m) \right]dz.
\end{eqnarray}

For spherically symmetric black holes, the expression for the time
delay can be advantageously simplified. Notice that, for
spherically symmetric spacetimes, $D=0$ and
\begin{equation}
\tilde P_1(x,x_0)|_{D=0}=\sqrt{\frac{B A_0}{A}}.
\end{equation}
Then the last integral in (\ref{T1T2b}) identically vanishes. When
$D \neq 0$, the dependence on $x_0$ remains through the impact
parameter $u$ which is present in Eq. (\ref{P1t}). The second
integral in (\ref{T1T2b}) can be approximated substituting the
integrand with $\sqrt{B_m/A_m}$ since it is practically constant
throughout the (very small) integration interval. Finally,
combining (\ref{Tt u}) with (\ref{S F L theta}) we get a very
simple expression for the first term in (\ref{T1T2b}).

\begin{figure}
\resizebox{\hsize}{!}{\includegraphics{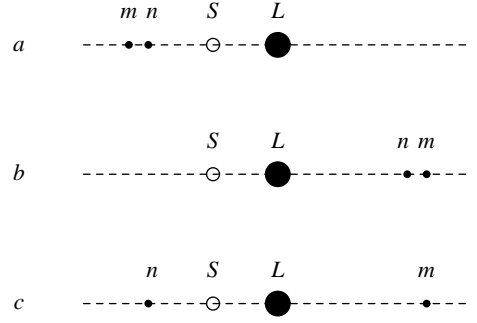}}
 \caption{This figure illustrates which images are considered in
 the different time delay calculations in the text. $L$ represents the lens,
 $S$ is the source. On each side of the lens an infinite series of images is formed.
 In case $a$, we consider two images on the same side of the
 source. Their time delay is given by $\Delta T^s_{n,m}$ with the
 upper sign.  In case $b$, we consider two images appearing on the same side
 but opposite to the source. Their time delay is given by $\Delta T^s_{n,m}$ with the
 lower sign. Finally, in case $c$, we consider two images appearing on opposite
 sides. Their time delay is given by $\Delta T^o_{n,m}$.
 }
 \label{Fig Conf}
\end{figure}

In writing the final formulae, we distinguish the case when the
two images are on the same side of the lens from the case when the
two images are on opposite sides of the lens. In the first case,
we have
\begin{eqnarray}
&\Delta & \!\!\!\! T^s_{n,m}= 2\pi (n-m)\frac{\tilde
a}{\overline{a}} \nonumber \\ &&+ 2\sqrt{\frac{B_m}{A_m}}
\sqrt{\frac{u_m}{c}} e^{\frac{\overline{b}}{2\overline{a}}} \left(
e^{-\frac{2m\pi\mp\gamma}{2\overline{a}}}-e^{-\frac{2n\pi
\mp\gamma }{2\overline{a}}} \right),\label{DTs}
\end{eqnarray}
where the upper sign before $\gamma$ applies if both images are on
the same side of the source (Fig. \ref{Fig Conf}a) and the lower
sign if both images are on the other side (Fig. \ref{Fig Conf}b).

If the images are on opposite sides of the lens (Fig. \ref{Fig
Conf}c), then
\begin{eqnarray}
&\Delta &  \!\!\!\! T^o_{n,m}= \left[ 2\pi (n-m) -2\gamma
\right]\frac{\tilde a}{\overline{a}} \nonumber \\ &&+
2\sqrt{\frac{B_m}{A_m}} \sqrt{\frac{u_m}{c}}
e^{\frac{\overline{b}}{2\overline{a}}} \left(
e^{-\frac{2m\pi-\gamma}{2\overline{a}}}-e^{-\frac{2n\pi +\gamma
}{2\overline{a}}} \right),\label{DTo}
\end{eqnarray}
where the image winding $n$ times is on the same side of the
source and the other is on the opposite side.

Notice that the geometry mostly favoured for the observation of
relativistic images is that with the source almost aligned with
the lens, so that $\gamma \sim D_{OL}^{-1} \ll 2\pi$. Therefore
\begin{equation}
\Delta T^o_{n,n} \ll  T^s_{n,m \neq n} \simeq  T^o_{n,m \neq n}
\end{equation}
i.e. if we evaluate the time delay between images with the same
winding number on opposite sides, we generally find a value which
is much smaller than the time delay between images with different
winding number.

Moreover, for physically reasonable values of the coefficients
$\overline{a}$, $\overline{b}$, which are all of order one, the
second term at the right hand side of Eq. (\ref{DTo}) is much
smaller than the first. For example, in the Schwarzschild black
hole, the time delay between the first and the second relativistic
images is (in Schwarzschild units)
\begin{equation}
\Delta T_{2,1}=16.57 \label{Schw}
\end{equation}
where the second term contributes only for $1.4\%$ to the total
time delay.

For spherically symmetric metrics, we also have the very important
relation
\begin{equation}
\frac{\tilde a}{\overline{a}}= u_m.
\end{equation}
Namely, the dominant term in the time delay is not a new
independent combination of the black hole metric function and
gives no further hint for the classification of the black hole. On
the contrary, the subdominant term is an independent combination
and could be used in principle to constrain the black hole model.
However, the subdominant term would be typically hidden below the
observational precision and it becomes reasonable to approximate
the time delay by its dominant contribution. In this way, a very
interesting surprise arises. In fact, suppose we are able to
measure the time delay between the first two images. Once we
restore physical units, the ratio between this time delay and the
minimum impact angle is
\begin{equation}
\frac{\Delta T_{2,1}}{\theta_m}=2\pi \frac{D_{OL}}{c_0},
\label{distance}
\end{equation}
where $c_0$ is the speed of light. In principle, by this formula,
we can get a very accurate estimate for the distance of the black
hole and hence of the whole hosting galaxy. The feasibility of
such an estimate, will be discussed in the next section.

In Appendix A, we treat the Kerr metric as an example of
non-spherically symmetric black hole. In that case, most of the
simplifications we have done, do not apply.

\section{Time delay in supermassive black hole lensing} \label{Sec Phy}
In order to achieve a complete reconstruction of the
characteristics of the black hole by strong field gravitational
lensing, we must distinguish at least the outermost relativistic
image from the others. Yet, as noted in \cite{Boz1}, in order to
achieve this, we need an optical resolution one or two orders of
magnitude better than that reachable by short-term VLBI projects
\cite{VLBI}. Therefore, relativistic images will possibly become a
target for next generation projects. With this in mind, we can
proceed to give estimates for time delays between the first and
the second relativistic images in realistic situations.

We treat only black holes with spherical symmetry, because only in
this case we have the formation at first of more than one
observable image. In fact, as noted in \cite{Boz2}, the
phenomenology of spinning black holes is quite different. In
particular, if the source is not inside a caustic, only one image
should become visible, while all the others stay very faint. On
the contrary, if the source is inside a caustic, two additional
non-equatorial images should appear. But an analytical treatment
for these additional images is not available at present.

Of course, we implicitly assume that the source must have temporal
variations, otherwise there is no time delay to measure. Thus, an
essential condition is that the source must be somehow variable.
However, this is not a so restrictive requirement, since variable
stars are generally abundant in all galaxies.

In Table \ref{S-LG-TD}, we present the values of the time delay
for the black hole located at the center of the Milky Way and in
other two galaxies of the Local Group. The results are obtained
using the Schwarzschild metric. It is clear that we have a little
chance to observe such short time delays for reasonable times of
exposure.
\begin{table}
\begin{tabular}{|l|c|c|c|}
  \hline
  Local Group  & Mass & Distance & Schwarzschild \\
 Galaxy &  $(\mathrm{M}_\odot)$ & (Mpc) & $\Delta T_{2,1}$ \\
    \hline
    Milky Way & $2.8\times10^6$ & 0.0085 & 0.1 h\\
    NGC0221 (M32)& $3.4\times10^6$ & 0.7 & 0.2 h\\
    NGC0224 (M31)& $3.0\times10^7$ & 0.7 & 1.4 h\\
    \hline
\end{tabular}
\caption{ \label{S-LG-TD} Estimates for the time delay for the
supermassive black hole located at the centers of three galaxies
in the case of Schwarzschild spacetime geometry. The masses and
the distances are taken from Richstone et al. \cite{Ric}.}
\end{table}

In order to have higher time delays, we need black holes with
larger Schwarzschild radii, i.e. more massive black holes. At the
same time we require that the magnification of the images must
remain of the same order. Since we know that the magnification is
proportional to $M_{\mathrm{Lens}}/D_{\mathrm{OL}}$, our request
can be fulfilled if we consider lenses with a mass of two or three
orders of magnitude larger than the black hole in the center of
our galaxy, and located not farther than three orders of magnitude
its distance. In this case, the measurement of the time delay
becomes more favorable as shown in Table \ref{S-G-TD}, where we
report our estimates for the time delay due to supermassive black
holes located at the centers of not too far galaxies, according to
spacetime geometry.
\begin{table}
\begin{tabular}{|l|c|c|c|}
  \hline
    & Mass & Distance & Schwarzschild \\
 Galaxy &  $(\mathrm{M}_\odot)$ & (Mpc) & $\Delta T_{2,1}$ \\
    \hline
    NGC4486 (M87) & $3.3\times10^9$ & 15.3 & 149.3 h\\
    NGC3115 & $2.0\times10^9$ & 8.4 & 90.5 h\\
    NGC4374 (M84)& $1.4\times10^9$ & 15.3 & 63.3  h\\
    NGC4594& $1.0\times10^9$ & 9.2 & 45.2 h\\
    NGC4486B (M104)& $5.7\times10^8$ & 15.3 & 25.8 h\\
    NGC4261 & $4.5\times10^8$ & 27.4 & 20.4 h\\
    NGC7052 & $3.3\times10^8$ & 58.7 & 14.9 h\\
    NGC4342 (IC3256)& $3.0\times10^8$ & 15.3 & 13.6 h\\
    NGC3377 & $1.8\times10^8$ & 9.9 & 8.1 h\\
\hline
\end{tabular}
\caption{ \label{S-G-TD} Estimates for the time delay for
supermassive black holes located at the center of several nearby
galaxies in the case of Schwarzschild spacetime geometry. The
masses and the distances of the central black holes are taken from
Richstone et al. \cite{Ric}.}
\end{table}

The time delays range from few hours to several days. It must be
kept in mind that a very deep exposure is needed to detect the
very faint relativistic images. The precise time will depend on
the characteristics of the future interferometers which will catch
the relativistic images and on the power of the source. However,
we can imagine that an exposure of 10 hours can be still taken as
a reasonable reference value for a deep imaging of a supermassive
black hole. Then, with a high enough sampling and a suitable
periodicity for the variable source, we can imagine to determine
the time delay with an accuracy of few hours. So, most of the
black holes in Tab. \ref{S-G-TD} would yield measurable time
delays.

Now consider the supermassive black hole in M87 and suppose we
manage to reach an accuracy of $5\%$ in a time delay measure. The
resolution needed to resolve the first two images is $0.01$
$\mu$arcsecs, while the minimum angle is $\theta_m=11$
$\mu$arcsecs. From formula (\ref{distance}), we can get the
distance to M87 with an accuracy of $5\%$ (the error in the angle
measurement is negligible). This is already better than standard
estimates by classical distance indicators \cite{Fouque}, whose
accuracy ranges from $10\%$ to $25\%$.

So, gravitational lensing in the strong field limit may become a
potentially competitive distance estimator in a not so far future.
This is a consequence of the fact that the time delay is a
dimensional variable and thus immediately leads to the measure of
a scale. In the strong field frame, it is proportional to the mass
of the black hole through the minimum impact parameter. However,
it happens that we can also measure the minimum impact angle
$\theta_m=u_m/D_{OL}$ directly, so that their ratio leaves us with
the distance to the lens. A time measurement can be done with a
high accuracy and has the advantage of being completely immune
from any unwanted bias or systematics, unlike the classical
estimates relying on luminosity measurements and typically highly
model-dependent assumptions.

A measurement as simple as this cannot be realized in weak field
gravitational lensing, because it requires an accurate modeling of
the gravitational potential. Moreover, the length scale it
measures is in general a more involved combination of all
geometrical distances ($D_{OL}$, $D_{LS}$, $D_{OS}$).

Another interesting possibility of strong field gravitational
lensing is the possibility of getting a mass estimate. By the
characteristics of the first two relativistic images, we can get
the coefficients $\overline{a}$ and $\overline{b}$, according to
the procedure described in \cite{Boz1}. They are generally
sufficient to identify the class of the specific black hole.
Afterwards, we can guess the theoretical $u_m$ in Schwarzschild
radii for the specific black hole model. Combining with the the
observed $\theta_m$ and with the $D_{OL}$ obtained by time delay,
we get the Schwarzschild radius and hence the mass of the black
hole. So, in principle, the time delay measurement would make the
strong field gravitational lensing completely autonomous from
external inputs coming from other methods.

One final consideration about the subdominant term neglected in
(\ref{distance}): if we simply identify the time delay with its
dominant contribution, we overestimate it by 1 or 2$\%$. However,
once we have identified the black hole class by the coefficients
$\overline{a}$ and $\overline{b}$, we can easily evaluate the
expected contribution of the subdominant term on the specific
black hole model and subtract it from the observed time delay. We
are then left with the pure dominant term and no more systematic
errors (however small) are present.

\section{Conclusions}

Gravitational lensing in the strong field limit may represent a
key tool for the investigation of supermassive black holes. In
principle, a complete characterization of the parameters of a
black hole can be achieved by the study of the images formed by
gravitational lensing of a background source. Technically, this
study requires resolutions one or two orders of magnitudes better
than actual VLBI projects, so that it stands as a possible
observational target for the next future.

In this work we have pointed out that photons contributing to
different strong field images take different times to reach the
observer. This time delay is of order of few seconds for the black
hole at the center of our Galaxy, but amounts to several days for
more massive black holes at the centers of nearby galaxies.

If the background source is characterized by an intrinsic
variability, it would then be possible to measure the time delay
between different strong field images, with the important
advantage of gaining a dimensional measurement for the scale of
the system. This measurement can be immediately used to get an
accurate distance determination for the observed black hole, not
affected by any kind of bias or model-dependent assumption.
Identifying the black hole class by the use of the other strong
field limit observables, we can also derive the mass of the black
hole in a completely independent way. This result encourages our
belief that gravitational lensing in the strong field limit stands
as an interesting (maybe powerful) method for the classification
of black holes and the determination of their characteristics.
Moreover, a new independent distance determination method is
always welcome in cosmological contexts.

\begin{acknowledgments}

The authors are grateful to Gaetano Scarpetta for helpful comments
on the manuscript. V.B. whishes to thank the theoretical division
of CERN and the Institute of Theoretical Physics of Z\"urich
University for their hospitality.

\end{acknowledgments}

\appendix{\section{Time delay in Kerr black holes}

If the black hole is not spherically symmetric, the
simplifications described at the end of Sect. \ref{Sec TD} do not
apply. In particular, the third integral in (\ref{T1T2b}) does not
identically vanish. In this appendix we work out the time delay in
a Kerr back hole as an example of non-spherically symmetric black
metric.

The Kerr metric projected on the equatorial plane reads
\begin{eqnarray}
&&A(x) = 1 - \frac{1}{x}  \\ &&B(x) =
\frac{1}{1-\frac{1}{x}+\frac{a^2}{x^2}} \\ &&C(x) =
x^2+a^2+\frac{a^2}{x} \\ &&D(x) = 2\frac{a}{x},
\end{eqnarray}
where $a$ is the specific angular momentum of the black hole.

We start directly from Eq. (\ref{T1T2b}), but we can still express
the dominant term $\Delta \tilde T$ in a simpler form. Consider
first the case of two images on the same side of the black hole.
Then everything works in the same way as for spherically symmetric
black holes and we get
\begin{equation}
\Delta \tilde T^s_{n,m}=2\pi(n-m) \frac{\tilde a}{\overline{a}}.
\end{equation}

Of course, the values of $\tilde a$ and $\overline{a}$ depend on
the sign of the spin $a$, i.e. they are different for photons
winding in the same sense of the black hole (direct photons) and
for photons winding in the opposite sense (retrograde photons).

If we wish to evaluate the time delay between two relativistic
images appearing on opposite sides of the black hole, then we have
to take care of the fact that one image will be direct and the
other will be retrograde. We then get
\begin{eqnarray}
&\Delta \tilde T^o_{n,m}&=\frac{\tilde a(a)}{\overline{a}(a)}
[2\pi n+\gamma- \overline{b}(a)]+ \tilde b(a)- \nonumber \\
&&\frac{\tilde a(-a)} {\overline{a}(-a)} [2\pi m-\gamma-
\overline{b}(-a)]-\tilde b(-a)
\end{eqnarray}
and we see that now we also need the coefficients $\overline{b}$
and $\tilde b$ for the calculation, since they are not the same
for the two images and do not cancel like in the spherically
symmetric case.

\begin{figure}
\resizebox{\hsize}{!}{\includegraphics{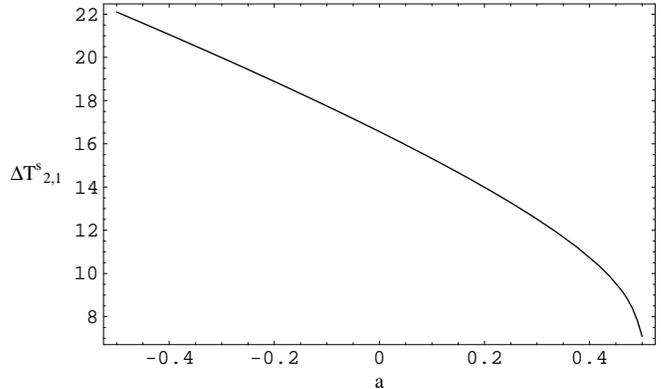}}
 \caption{Time delay (as a function of the black hole spin)
between the second and the first relativistic images appearing on
the same side of a Kerr black hole.}
 \label{Fig Kerr1}
\end{figure}

Considering a source aligned behind the black hole ($\gamma=0$),
in Fig. \ref{Fig Kerr1} we plot the time delay between the second
and first images appearing on the same side of the black hole. For
positive $a$ the two images are direct and for negative $a$ they
are both retrograde. We see that the time delay decreases if the
images are both direct, while increases if they are both
retrograde. We can also notice that the largest contribution to
the time delay still comes from $\Delta \tilde T$, while the
second term in (\ref{T1T2b}) at most contributes for $6\%$ when
$a=0.5$ and the last term stays below $0.7\%$.

\begin{figure}
\resizebox{\hsize}{!}{\includegraphics{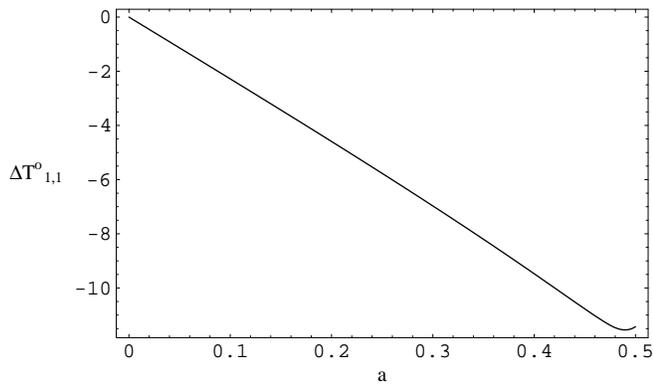}}
 \caption{Time delay (as a function of the black hole spin)
between the first direct relativistic image and the first
retrograde image.}
 \label{Fig Kerr2}
\end{figure}

The situation is quite different for images on opposite sides
(Fig. \ref{Fig Kerr2}). The time delay is zero when $a=0$ and
becomes negative for positive $a$. This means that direct light
rays take less time than retrograde rays to wind around the black
hole. This is naturally understood since the radius of the photon
sphere is larger for retrograde light rays. For high values of the
black hole spin, $\Delta T^o_{1,1}$ becomes comparable to $\Delta
T^s_{2,1}$. Another interesting fact is that the second term in
(\ref{T1T2b}) is of the same order of the dominant term $\Delta
\tilde T$, the ratio being roughly $-1/3$. The last term also
contributes for $1.7\%$. }


\begin{thebibliography}{02}

\bibitem{DED} F.W. Dyson, A.S. Eddington, C. Davidson, Phil. Trans. Roy. Soc. 220A, 291 (1920).

\bibitem{SEF} P. Schneider, J. Ehlers, E.E. Falco,
\textit{Gravitational lenses}, Springer-Verlag, Berlin (1992).

\bibitem{Atk}  R.D. Atkinson, Astron. Jour. {\bf 70}, 517 (1965).

\bibitem{VirEll}  K.S. Virbhadra, G.F.R. Ellis, Phys. Rev. D {\bf 62}, 084003 (2000).

\bibitem{Vie}  S.U. Viergutz, A\&A 272 (1993) 355.

\bibitem{Nem} R.J. Nemiroff, Amer. Jour. Phys. {\bf 61}, 619 (1993).

\bibitem{FN} S. Frittelli, E.T. Newman, Phys.
Rev. D {\bf 59}, 124001 (1999).

\bibitem{FKN} S. Frittelli, T.P. Kling, E.T. Newman, Phys.
Rev. D {\bf 61}, 064021 (2000).

\bibitem{Per}  V. Perlick, gr-qc/0307072.

\bibitem{DabSch}  M.P. Dabrowski, F.E. Schunck, Astoph. Jour. {\bf 535}, 316 (2000).

\bibitem{FMA}  H. Falcke, F. Melia, E. Agol, ApJ Letters 528 (1999)
L13.

\bibitem{BCIS} V. Bozza, S. Capozziello, G. Iovane, G.
Scarpetta, Gen. Rel. and Grav. {\bf 33}, 1535 (2001).

\bibitem{ERT} E.F. Eiroa, G.E. Romero, D.F. Torres, Phys. Rev. D {\bf 66},
024010 (2002).

\bibitem{Pet} A.O. Petters, MNRAS {\bf 338}, 457 (2003).

\bibitem{Boz1} V. Bozza, Phys. Rev. D {\bf66}, 103001 (2002).

\bibitem{Bha} A. Bhadra, Phys. Rev. D {\bf 67}, 103009 (2003).

\bibitem{Boz2} V. Bozza, Phys. Rev. D {\bf 67}, 103006 (2003).

\bibitem{VazEst} S.E. Vazquez, E.P. Esteban, gr-qc/0308023.

\bibitem{Ref} S. Refsdal, MNRAS {\bf 128}, 307 (1964).

\bibitem{BlaNar} R.D. Blandford, R. Narayan, Ann. Rev. Astron. \&
Astroph. {\bf 30}, 311 (1992).

\bibitem{WCW} D. Walsh, R.F. Carswell, R.J. Weymann, Nature {\bf 279}, 381 (1979).

\bibitem{VLBI} ARISE web page: arise.jpl.nasa.gov; MAXIM web page:
maxim.gsfc.nasa.gov; J.S. Ulvestad astro-ph/9901374.

\bibitem{Ric} D. Richstone et al., Nature {\bf 395}, A14 (1998).

\bibitem{Fouque} P. Fouqu\'e, J.M. Solanes, T. Sanchis, C. Balkowski, Astron. and Astroph.
{\bf 375}, 770 (2001).

\end{thebibliography}
\end{document}